\documentclass[11pt,nofootinbib,showpacs,notitlepage]{revtex4-1}
\usepackage[T1]{fontenc}
\usepackage{babel}
\usepackage[utf8]{inputenc}
\usepackage{babel,amsmath,amssymb,mathrsfs,amsfonts,calligra,euscript,textcomp,graphicx,wrapfig,fontenc,pst-blur,tabu,lipsum,%
float,empheq,pst-node,environ,framed,ccaption,capt-of,enumitem,subcaption,enumitem,framed,txfonts,lmodern,cancel,diagbox,multirow,float,tabularx}
\usepackage[section]{placeins}
\definecolor{darkgreen}{rgb}{0,.7,0}
\usepackage[colorlinks,linkcolor=blue,citecolor=green,pagebackref=true]{hyperref}
\def\[{\left[}
\def\]{\right]}
\def\({\left(}
\def\){\right)}

\newcommand{\const}{\mathop{\rm const}\nolimits}

\newcommand{\m}{\mathop{\mathcal{M}}\nolimits}

\newcommand{\eq}[1]{\begin{equation}#1\end{equation}}

\begin{document}
\title{Spontaneous symmetry breaking as a result of extra dimensions compactification}

\author{Dmitry Chirkov}
\affiliation{Sternberg Astronomical Institute, Moscow State University, Moscow, Russia}

\author {Alex Giacomini}
\affiliation{Instituto de Ciencias Físicas y Matemáticas, Universidad Austral de Chile, Valdivia, Chile}

\author{Alexey Toporensky}
\affiliation{Sternberg Astronomical Institute, Moscow State University, Moscow, Russia}
\affiliation{Kazan Federal University, Kremlevskaya 18, Kazan 420008, Russia}

\author {Petr Tretyakov}
\affiliation {Joined Institute for Nuclear Research, Dubna, Russia}

\begin{abstract}
   We consider dynamics of a scalar field in compactification scenario of Einstein-Gauss-Bonnet cosmology. It is shown that if the field is non-minimally coupled to curvature, its asymptotic value under certain conditions may be shifted from the minimum of its potential. This means that due to
influence of extra dimensions a scalar field with $\lambda \phi^4$ potential can stabilise away from $\phi=0$ stable point which means an effective symmetry breaking occurs in such a system.
\end{abstract}

\maketitle

\section{Introduction}
The idea of modifying gravity on a cosmological scale has become widespread in the last few decades. This has been triggered both by theoretical developments and observations. Since the first days after Einstein’s publication of his theory there were proposals being made on how to incorporate it in a more unified theory. Examples of this are Eddington’s theory of connections~\cite{Eddington}, Weyl’s scale independent theory~\cite{Weyl}, the higher dimensional theories of Kaluza and Klein~\cite{Kaluza,Klein}. Later building on Weyl’s works Sakharov proposed that the Einstein-Hilbert action is just a first approximation to a much more complicated action~\cite{Sakharov}. Stelle showed that theories with a higher power corrections are renormalizable in the presence of matter fields at the one loop level~\cite{Stelle-1,Stelle-2}.  This discovery was followed by a great interest to the potential cosmological consequences of these theories~\cite{Starobinsky}. On the other hand, the limits of General Relativity on cosmological scales have come into focus with the appearance of the "dark universe" scenario (in order to fit the astrophysical observations one must assume the existence of dark matter and dark energy). Another issues is that due to the fact that the Standard Model of particle physics is based on perturbative quantum field theory, gravity does not fit into it (a naive attempt to quantize gravity leads to a non-renormalizable theory).

Currently, there are a huge number of modified theories of gravity in the literature (see~\cite{Clifton} for comprehensive overview). Some of these have extra scalar, vector or tensor fields in their gravitational sector; some develop Sakharov's idea by modifying gravity in regions of low rather than high curvature; others expand on the ideas put forward by Kaluza and Klein.  In the context of higher dimensional gravity a very natural choice for a modified theory of gravity is given by Lovelock gravity~\cite{Lovelock}. Lovelock models are characterized by the fact that their actions possess higher power curvature terms but whose variation leads to equations of motion which remain of second order derivative in the metric.

The most studied particular case of Lovelock model is the Einstein-Gauss-Bonnet gravity. The Lagrangian of this theory is the sum of the Einstein-Hilbert term and so-called Gauss-Bonnet term $R^2-4R_{\mu\nu}R^{\mu\nu}+R_{\mu\nu\zeta\eta}R^{\mu\nu\zeta\eta}$; for (3+1)-dimensional space-time the Gauss-Bonnet term is topological and does not affect the dynamical equations; in dimensions higher than four this term gives a non-trivial contribution to equation of motion.

Cosmology in Lovelock gravity and, particularly in the Einstein-Gauss-Bonnet (EGB) gravity have been studied recently rather intensively~\cite{Müller-Hoissen-1985,Madore-1985,Madore-1986,Müller-Hoissen-1986,DFB,Verwimp,DCMTP,MM,CGTW,Ivashchuk-1,grg10,KPT,Iv-09-1,Iv-09-2,Odin1,Odin2,CGPT1,CGPT2,CST1,ChPavTop2,ChPavTop1,Pavl-15,ErIvKob-16,Iv-16,ChT,Pavl,Fr,CGT,CT:splitting,ChPavl}. Quite interesting results have been obtained including cosmological regimes which are impossible in General relativity. While power-law vacuum solutions in EGB cosmology resembles known Kasner solution of GR~\cite{Deruelle,DFB}, a non-zero cosmological constant changes the situation drastically. In GR positive $\Lambda$-term ultimately leads to de Sitter solution, in EGB gravity in addition to de Sitter solution the other anisotropic type of solutions with constant but different Hubble parameters $H_i$ appears~\cite{Ivashchuk-1,grg10,KPT}. However, these Hubble parameters can not be totally different, the maximum number of different $H_i$ can not exceed $3$ regardless the number of dimensions~\cite{Iv-16}. Most of such solutions are stable (see for details~\cite{Iv-16,Pavl-15,ErIvKob-16,ChT,ChPavl}) and numerical integrations show that they represent a typical attractor for a flat multidimensional cosmological dynamics. This means that initially totally anisotropic Universe tends to form a product of two or three isotropic subspaces. From the perspective of dynamical compactification the situation of two subspaces, one of which is expanding and the other is contracting, is of a particular interest. It is not difficult to choose the coupling constant of the theory so that this requirement is fulfilled, particular examples and numerical confirmation have been done in~\cite{CT:splitting}.

So that setting the multidimensional space metric to be a product of two isotropic subspaces in the framework of EGB gravity is justified.  This form is much easier to study analytically and generalize to non-zero spatial curvature. The latter is important since it is the curvature of extra dimension space that is responsible to
extra space stabilizing~\cite{CGTW}. So that, a scenario where initially anisotropic space splits dynamically into product of two isotropic subspaces, and later "inner"
subspace stabilizes seems to be rather general though its full treatment is still to be done (possible influence of curvature on the first stage needs a particular
investigation). Once inner dimensions stabilize, the effective dynamics of the bigger subspace is essentially a Friedmann dynamics~\cite{Fr}. This is correct for a vacuum Universe, Universe with a cosmological constant or Universe filled with a barotropic fluid. The goal of the present paper is to study the cosmological dynamics in the case when Universe is filled with a scalar field. We will see that if the scalar field is non-minimally coupled with curvature, extra terms appear in the effective dynamics of
the bigger subspace after stabilization of the inner subspace.

\section{Action and equations of motion}

We consider 8-dimensional spacetime $\mathcal{M}=\mathcal{L}_4\times \m_4$ where $\mathcal{L}_4$ is a flat Friedman-Robertson-Walker manifold with scale factor $a(t)$, $\m_4$ is a 4-dimensional Euclidean compact constant curvature manifold with scale factor $b(t)$ and negative spatial curvature. We take metric to be of the form

\eq{ds^2=-dt^2+a(t)^2\left(dx^2+dy^2+dz^2\right)+b(t)^2\left[d\psi^2+\sinh^2\psi\left(d\chi^2+\sin^2\chi d\theta^2+\sin^2\chi\sin^2\theta d\phi^2\right)\right]\label{metric}}
Action under consideration reads

\eq{S=\int_{\mathcal{M}}d^{8}x\sqrt{|g|}\left\{\left(m_{\rm Pl}^2+\xi\phi^2\right)R-2\Lambda+\alpha L_{GB}-\frac{1}{2}g^{\rho\eta}\nabla_{\rho}\phi\nabla_{\eta}\phi-V(\phi)\right\},\label{action}}
where $m_{\rm Pl}$ is the 8-dimensional Planck mass, $g$ is the determinant of metric tensor; $\phi$ is a spatially homogeneous scalar field with the potential  $V(\phi)$; $\Lambda$ is a bare cosmological constant; $\alpha$ and $\xi$ are the coupling constants; $L_{GB}$ is quadratic Lovelock term:

\eq{L_{GB}=R^2-4R_{\mu\nu}R^{\mu\nu}+R_{\mu\nu\zeta\eta}R^{\mu\nu\zeta\eta}}
where $R,R_{\mu\nu},R_{\mu\nu\zeta\eta}$ are the $8$-dimensional scalar curvature, Ricci tensor and Riemann tensor, respectively\footnote{Hereafter Greek indices run from 0 to 7, while Latin one from 1 to 7 unless otherwise stated}.

Equations of motion that follow from the action take the form

\eq{\ddot{\phi}+\frac{\dot{g}}{2g}\dot{\phi}+V'-2\xi\phi R=0\label{eq:scalar_field}}

\eq{
\begin{array}{c}
    \left(m_{\rm Pl}^2+\xi\phi^2\right)G^{\mu}_{\phantom{\mu}\nu}+\alpha E^{\mu}_{\phantom{\mu}\nu}+ \vspace{0.2cm}\\
    +g_{\nu\sigma}\Biggl\{2\xi\left[\phi\dot{\phi}\left(2\cfrac{d}{dt}\cfrac{\partial R\sqrt{|g|}}{\partial\ddot{g}_{\sigma\mu}}-\cfrac{\partial R\sqrt{|g|}}{\partial\dot{g}_{\sigma\mu}}\right)+\left(\dot{\phi}^2+\phi\ddot{\phi}\right)\cfrac{\partial R\sqrt{|g|}}{\partial\ddot{g}_{\sigma\mu}}\right]- \vspace{0.2cm}\\
    -\cfrac{\partial}{\partial{g}_{\sigma\mu}}\left[\left(2\Lambda+\frac{1}{2}g^{\rho\eta}\nabla_{\rho}\phi\nabla_{\eta}+V\right)\sqrt{|g|}\right]\Biggr\}=0
\end{array}\label{general_eqs_of_motion}}
where
\eq{G^{\mu}_{\phantom{\mu}\nu}=R^{\mu}_{\phantom{\mu}\nu}-\frac{1}{2}R\delta^{\mu}_{\phantom{\mu}\nu}}
and
\eq{E^{\mu}_{\phantom{\mu}\nu}=2\left(R^{\mu}_{\phantom{\mu}\gamma\zeta\eta}R_{\nu}^{\phantom{\mu}\gamma\zeta\eta}-2R^{\mu}_{\phantom{\mu}\gamma\nu\eta}R^{\gamma\eta}
-2R^{\mu}_{\phantom{\mu}\gamma}R_{\nu}^{\phantom{\nu}\gamma}+RR^{\mu}_{\phantom{\mu}\nu}\right)-\frac{1}{2}L_{GB}\delta^{\mu}_{\phantom{\mu}\nu}}.

Substituting~(\ref{metric}) into~(\ref{eq:scalar_field}) and~(\ref{general_eqs_of_motion}) we get
\eq{{\ddot{\phi}}+\left(3 H +\frac{4 {\dot{b}}}{b }\right) {\dot{\phi}}+V'-2 \xi  \phi  \left(12 H ^{2}+6 {\dot{H}}+\frac{24 {\dot{b}} H }{b }+\frac{8 {\ddot{b}}}{b }+\frac{12 {\dot{b}}^{2}}{b ^{2}}-\frac{12}{b ^{2}}\right)=0\label{eq:scalar_field-bH}}
where prime stands for derivative with respect to $\phi$,


\eq{\begin{split}
      & \left(-\frac{6}{b^{2}}+3 H^{2}+2 {\dot{H}}+\frac{4 {\ddot{b}}}{b}+\frac{6 {\dot{b}}^{2}}{b^{2}}+\frac{8 {\dot{b}} H}{b}\right)\left[m_{\mathit{Pl}}^{2}+\phi^{2}\xi\right] +4\xi\phi\dot{\phi}\left(H+\frac{2{\dot{b}}}{b}\right)+2\xi\left(\dot{\phi}^2+\phi\ddot{\phi}\right)+\\
      & +\alpha\Biggl(\frac{12 {\dot{b}}^{4}}{b^{4}}-\frac{48 {\ddot{b}}}{b^{3}}-\frac{24 {\dot{b}}^{2}}{b^{4}}+\frac{12}{b^{4}}-\frac{48 \left({\dot{H}}+H^{2}\right)}{b^{2}}-\frac{24 H^{2}}{b^{2}}+\frac{48 {\ddot{b}} {\dot{b}}^{2}}{b^{3}}-\frac{96 {\dot{b}} H}{b^{3}}+\frac{48 \left({\dot{H}}+H^{2}\right) {\dot{b}}^{2}}{b^{2}}+ \\
      & \hspace{1cm}+\frac{16 {\ddot{b}} H^{2}}{b}+\frac{72 {\dot{b}}^{2} H^{2}}{b^{2}}+\frac{96 H {\dot{b}}^{3}}{b^{3}}+\frac{32 \left({\dot{H}}+H^{2}\right) H {\dot{b}}}{b}+\frac{96 {\ddot{b}} H {\dot{b}}}{b^{2}}\Biggr)=-\frac{{\dot{\phi}}^{2}}{4}+\frac{V}{2}+\Lambda,
    \end{split}\label{eq:1-1-bH}}

\eq{\begin{split}
      &\Biggl(6 H^{2}+3 {\dot{H}}+\frac{3 {\ddot{b}}}{b}+\frac{3 {\dot{b}}^{2}}{b^{2}}-\frac{3}{b^{2}}+\frac{9 {\dot{b}} H}{b}\Biggr)\left[m_{\mathit{Pl}}^{2}+\phi^{2}\xi\right]+6\xi\phi\dot{\phi}\left(H+\frac{{\dot{b}}}{b}\right)+2\xi\left(\dot{\phi}^2+\phi\ddot{\phi}\right)+\\
      & +\alpha\Biggl(-\frac{36 {\dot{H}}}{b^{2}}-\frac{72 H^{2}}{b^{2}}+\frac{12 {\ddot{b}} {\dot{b}}^{2}}{b^{3}}+12 H^{2} \left({\dot{H}}+H^{2}\right)-\frac{12 {\ddot{b}}}{b^{3}}+\frac{36 {\ddot{b}} H^{2}}{b}+\frac{36 H^{3} {\dot{b}}}{b}+\frac{108 {\dot{b}}^{2} H^{2}}{b^{2}}+\frac{36 H {\dot{b}}^{3}}{b^{3}}- \\
      & \hspace{1cm}-\frac{36 {\dot{b}} H}{b^{3}}+\frac{36 \left({\dot{H}}+H^{2}\right) {\dot{b}}^{2}}{b^{2}}+\frac{72 \left({\dot{H}}+H^{2}\right) H {\dot{b}}}{b}+\frac{72 {\ddot{b}} H {\dot{b}}}{b^{2}}\Biggr) =-\frac{{\dot{\phi}}^{2}}{4}+\frac{V}{2}+\Lambda,
    \end{split}\label{eq:7-7-bH}}

\eq{\begin{split}
      & \left(3 H^{2}+\frac{12 {\dot{b}}H}{b}+\frac{6 {\dot{b}}^{2}}{b^{2}}-\frac{6}{b^{2}}\right)\left[m_{\mathit{Pl}}^{2}+\phi^{2}\xi\right]+2\xi\phi\dot{\phi}\left(3H+\frac{4{\dot{b}}}{b}\right)+\\
      & +\alpha\Biggl(\frac{48 H^{3} {\dot{b}}}{b}+\frac{216 {\dot{b}}^{2} H^{2}}{b^{2}}+\frac{144 H {\dot{b}}^{3}}{b^{3}}+\frac{12 {\dot{b}}^{4}}{b^{4}}-\frac{72 H^{2}}{b^{2}}
      -\frac{144 {\dot{b}} H}{b^{3}}-\frac{24 {\dot{b}}^{2}}{b^{4}}+\frac{12}{b^{4}}\Biggr)=\Lambda +\frac{{\dot{\phi}}^{2}}{4}+\frac{V}{2}.
    \end{split}\label{eq:0-0-bH}}

Since metric contains only two independent functions $a(t)$ and $b(t)$, we have two independent equations~(\ref{eq:1-1-bH}) and~(\ref{eq:7-7-bH}) as well as constraint~(\ref{eq:0-0-bH}).

\section{Numerical calculations}

In what follows we deal with $V=\lambda\phi^4$.

Compactification scenario implies that $\dot{H},\dot{b},\ddot{b}\underset{t\rightarrow\infty}{\longrightarrow}0$; $b(t)\underset{t\rightarrow\infty}{\longrightarrow}\nolinebreak b_a$, $H(t)\underset{t\rightarrow\infty}{\longrightarrow}\nolinebreak H_a$, where $b_a=\const$ and $H_a=\const$ are asymptotic values of the scale factor $b(t)$ and the Hubble parameter $H(t)$.

Let $\phi_a$ be an asymptotic of scalar field $\phi$ after compactification; substituting $\dot{H}=\dot{b}=\ddot{b}=\dot{\phi}=\ddot{\phi}=0,\;b=b_a,\;H=H_a,\;\phi=\phi_a$ into equations~(\ref{eq:scalar_field-bH})-(\ref{eq:0-0-bH}), we get \emph{asymptotic equations}:

\eq{4\lambda\phi_a^3-24\xi\left(H_a^{2}-\frac{1}{b_a^{2}}\right)\phi_a=0
\quad\Longleftrightarrow\quad \phi_a=0\quad\vee\quad\phi_a^2=\frac{6\xi}{\lambda}\left(H_a^{2}-\frac{1}{b_a^{2}}\right)\label{eq-phi-asympt-2}}

\eq{\left(3H_a^{2}-\frac{6}{b_a^{2}}\right)\left(m_{\rm Pl}^2+\xi\phi_a^2\right)+\alpha\left(\frac{1}{b_a^{2}}-6H_a^{2}\right)\frac{12}{b_a^{2}}=\Lambda+\frac{\lambda\phi_a^4}{2}\label{eq-1-asympt-2}}

\eq{\left(6H_a^{2}-\frac{3}{b_a^{2}}\right)\left(m_{\rm Pl}^2+\xi\phi_a^2\right)+\alpha \left(H_a^2-\frac{6}{b_a^{2}}\right)12H_a^2=\Lambda+\frac{\lambda\phi_a^4}{2}\label{eq-2-asympt-2}}

It is easy to see that asymptotic Klein-Gordon equation~(\ref{eq-phi-asympt-2}) has non-trivial solutions for $H_a>\frac{1}{b_a}$ if $\xi$ is positive and for $H_a<\frac{1}{b_a}$ if $\xi$ is negative.

In order to check stability of analytic solutions we use numerical integration of equations of motion. We vary the parameters
 $|\xi|$ and $\lambda$ run from $10^{-12}$ to $10^3$,  $b_a$ runs from 1 to $10^5$ in Planck units, $H_a$ is taken to be less than Planck unit, $\phi_a$ is evaluated from~(\ref{eq-phi-asympt-2}) according to the table~\ref{table.phia}:

\begin{table}[!h]

\begin{center}
\caption{Asymptotic values of the scalar field after compactification}
\label{table.phia}

  \begin{tabular}{|c|c|c|}

    \hline


    & $\xi>0$ & $\xi<0$ \\

    \hline

    $H_a>\frac{1}{b_a}$ & $\phi_a=\sqrt{\frac{6\xi}{\lambda}\left(H_a^{2}-\frac{1}{b_a^{2}}\right)}$ & $\phi_a=0$ \\

    \hline

    $H_a<\frac{1}{b_a}$ & $\phi_a=0$ & $\phi_a=\sqrt{\frac{6\xi}{\lambda}\left(H_a^{2}-\frac{1}{b_a^{2}}\right)}$ \\

    \hline

  \end{tabular}

\end{center}

\end{table}

\noindent Once $|\xi|,\;\lambda,\;b_a,\;H_a,\;\phi_a$ are fixed, we evaluate $\alpha$ and $\Lambda$ from~(\ref{eq-1-asympt-2})-(\ref{eq-2-asympt-2}):

\eq{\alpha=-\frac{b_{a}^{2}\left(\xi\phi_{a}^{2}+m_{\rm{Pl}}^{2}\right)}{4(H_{a}^{2} b_{a}^{2}-1)}\label{alpha}}

\eq{\Lambda=\frac{\left(6\left(\xi\phi_{a}^{2}+m_{\rm{Pl}}^{2}\right)H_{a}^{2}-\lambda\phi_a^4\right)H_{a}^{2}b_{a}^{4}+
\left(18\left(\xi\phi_{a}^{2}+m_{\rm{Pl}}^{2}\right)H_{a}^{2}+\lambda\phi_a^4\right)b_{a}^{2}+6\left(\xi\phi_{a}^{2}+m_{\rm{Pl}}^{2}\right)}{2(H_{a}^{2} b_{a}^{2}-1) b_{a}^{2}}\label{Lambda}}

The advantage to setting a solution $b_0$ and $H_0$ and then calculating the necessary values of $\alpha$ and $\Lambda$ is that it is possible to get the expressions in a reasonably simple form.

After that we specify initial values of dynamical variables. Normally, since we are interested in stability of the compactification solution, we choose initial values in the vicinity of asymptotic values $b_0\in(0.9b_a;1.1b_a)$, $H_0\in(0.9H_a;1.1H_a)$, $\phi_0\in(0.9\phi_a;1.1\phi_a)$, $\phi'_0\in(-0.01;0.01)$; but our experiments show that extension of range of initial values (for instance, choosing $\phi_0\sim10^6\phi_a$) does not affect qualitatively the results.

Besides compactified solutions there exist isotropic solutions which are defined by the equation
\eq{420\alpha H^4 + 21H^2m^2_{\rm Pl} - \Lambda = 0}
The table~\ref{table.isotropic} below outlines possible isotropic solutions and conditions for their existence.

\begin{table}[!h]
\begin{center}
\caption{Possible forms of isotopic solution}
\label{table.isotropic}
  \begin{tabular}{|c|c|c|}
    \hline
    & $\alpha<0$ & $\alpha>0$ \\
    \hline
    $\Lambda<0$ & $H^2=-\frac{21 m_{\rm{Pl}}^{2}+\sqrt{441 m_{\rm{Pl}}^{4}+1680 \alpha  \Lambda}}{840 \alpha}$ & No solutions \\
    \hline
    $\Lambda>0$ & $\begin{array}{c}
         H^2=\frac{-21 m_{\rm{Pl}}^{2}\pm\sqrt{441 m_{\rm{Pl}}^{4}+1680 \alpha  \Lambda}}{840 \alpha}  \\
         \mbox{for}\;\;\alpha\Lambda>-\frac{21 m_{\rm{Pl}}^{4}}{80}
    \end{array}$ & $H^2=\frac{-21 m_{\rm{Pl}}^{2}+\sqrt{441 m_{\rm{Pl}}^{4}+1680 \alpha  \Lambda}}{840 \alpha}$ \\
    \hline
  \end{tabular}
\end{center}
\end{table}

 \begin{figure}[!h]
   \begin{subfigure}[t]{.4\textwidth}
     \centering
     \includegraphics[width=\linewidth]{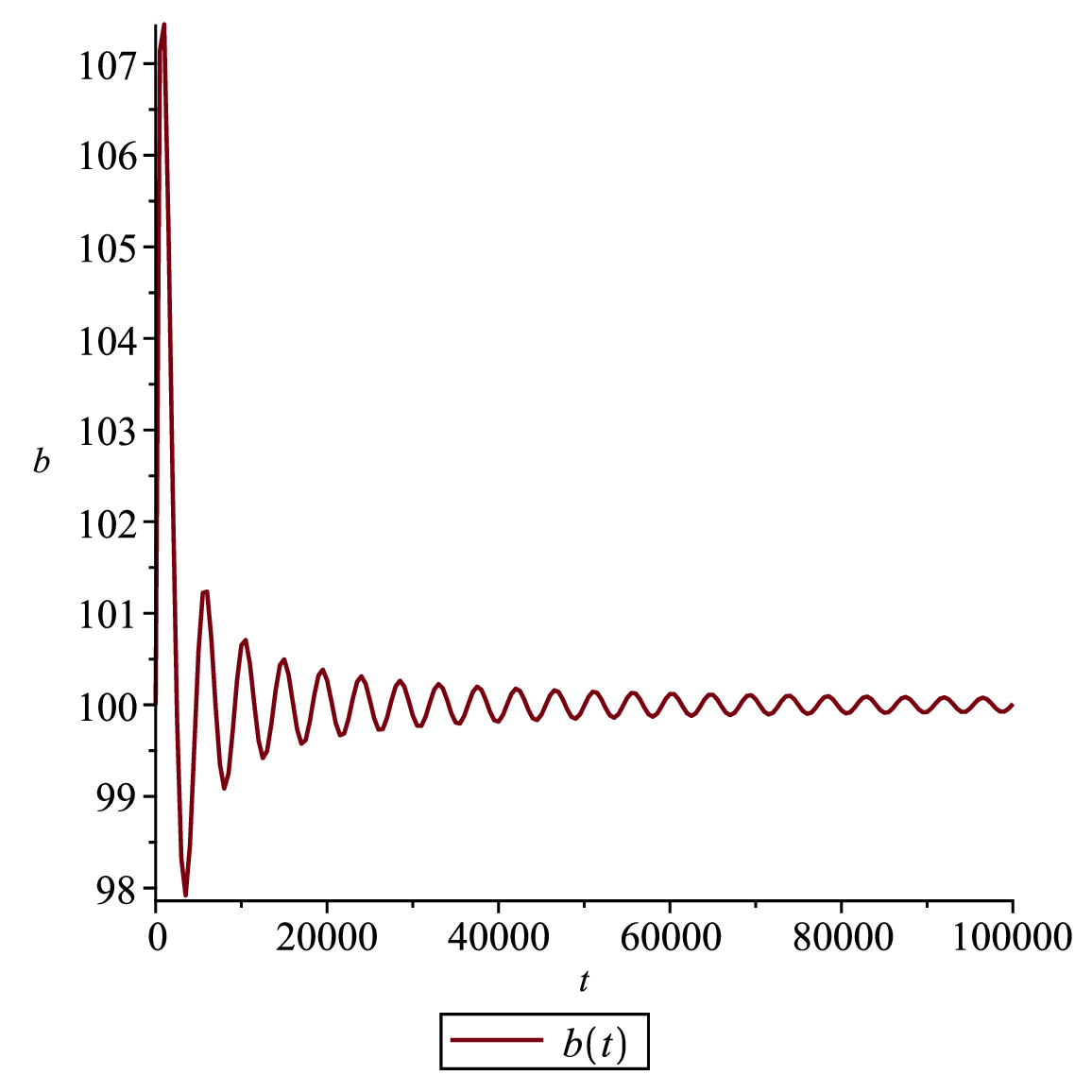}
   \end{subfigure}
   \hfill
   \begin{subfigure}[t]{.4\textwidth}
     \centering
     \includegraphics[width=\linewidth]{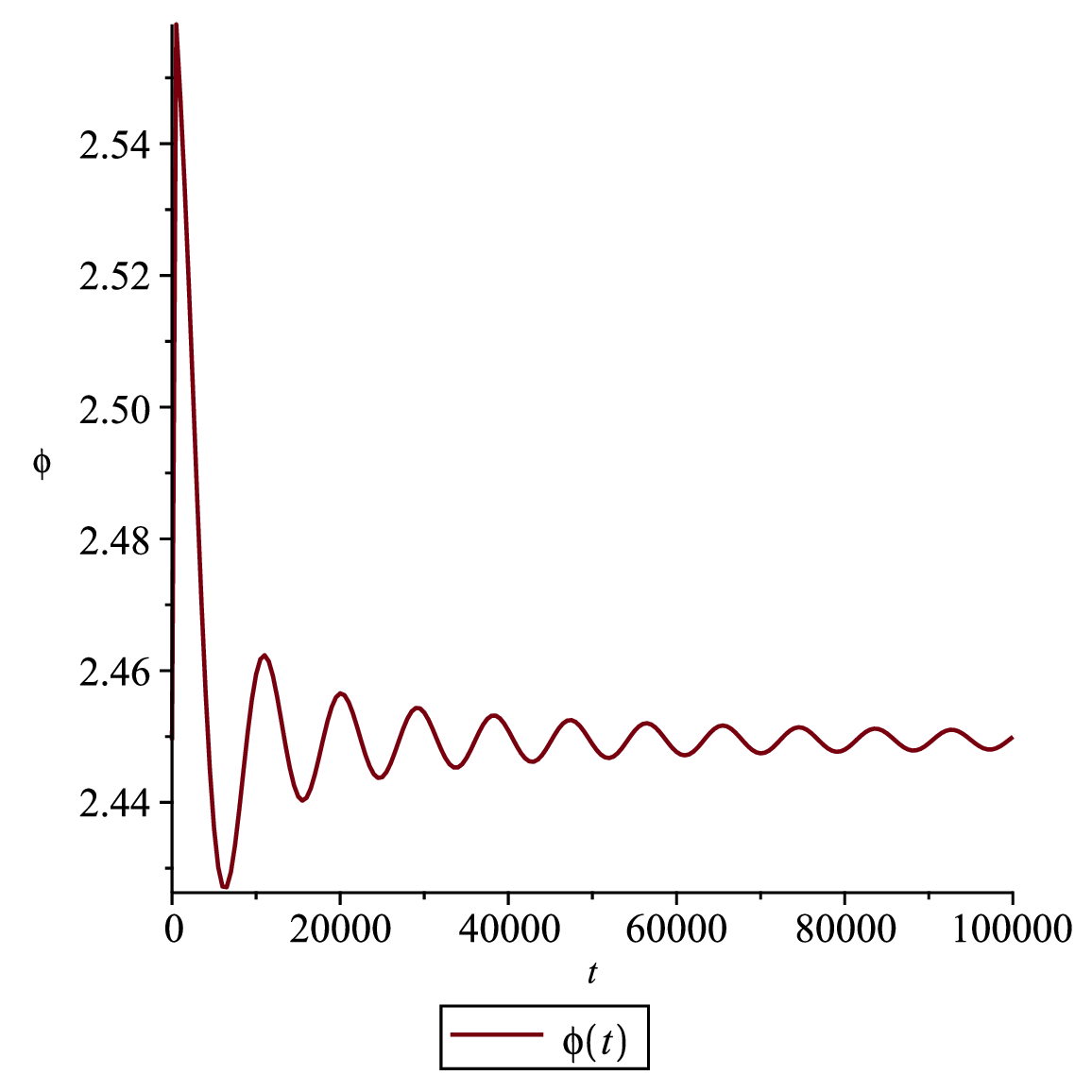}
   \end{subfigure}
   \caption{\footnotesize Typical compactified solution for $\alpha=2498.5,\; \Lambda=-0.0003,\; \xi=-10^{-4},\; \lambda=10^{-8}$:  a) behaviour of the scale factor $b(t)$; b) behaviour of the scalar field }
   \label{compactification}
 \end{figure}

Initial value $b'_0$ of derivative of the scale factor $b$ is found by solving constraint~(\ref{eq:0-0-bH}). It is a quartic equation with respect to $b'_0$ and it has up to 4 real roots.
Depending on coupling constant and initial $b'_0$ we choose we get singular solution, isotropic solution or compactified solution. Finding solution numerically usually  means its stability with respect to small homogeneous perturbations.
We have detected numerically compactified solutions in a wide range of parameter values $|\xi|,\;\lambda,\;b_a,\;H_a$, so a fine-tuning does not needed. We leave a detailed analysis of the range of stability for compactification solutions  for a future work.


\section{Zero cosmic acceleration case}

Realistic compactification regime assumes that the asymptotic value of the Hubble parameter $H(t)$ is extremely small in natural units. So that, substituting  $H(t)=0$ and $b(t)=b_a$ we get that the evolution of the scalar field is governed by an effective potential having this simple form

\eq{V_{\rm eff_{a}}=\lambda\phi^4+\frac{12\xi}{b_a^2}\phi^2}.

The point of minimum $\phi_{\rm min}$ is solution to $V'_{\rm eff_{a}}=0$ equation. A non-zero solution exists for negative values of $\xi$ only:

\eq{\phi_{\rm min}=\frac{1}{b_a}\sqrt{\frac{6|\xi|}{\lambda}}\label{phi-min}}
So that, we got an effective  Mexican hat potential for $\xi<0$ starting from a simple quartic bare potential.

In order to use this situation to generate a realistic Higgs potential we need that this value of the scalar field is small in Planck units. This can be obtained either for large $b$ or small $\xi$.

The formulae (\ref{alpha})-(\ref{Lambda}) in the $H_a=0$ case simplify as
\eq{\alpha=\frac{b_{a}^{2} \left(\xi\phi_{a}^{2}+m_{\rm{Pl}}^{2}\right)}{4}}
\eq{\Lambda=-\frac{\lambda b_{a}^{2}\phi_{a}^{4}+6\xi\phi_{a}^{2}+6 m_{\rm{Pl}}^{2}}{2 b_{a}^{2}}}

\begin{figure}
  \begin{subfigure}[t]{.49\textwidth}
    \centering
    \includegraphics[width=\linewidth]{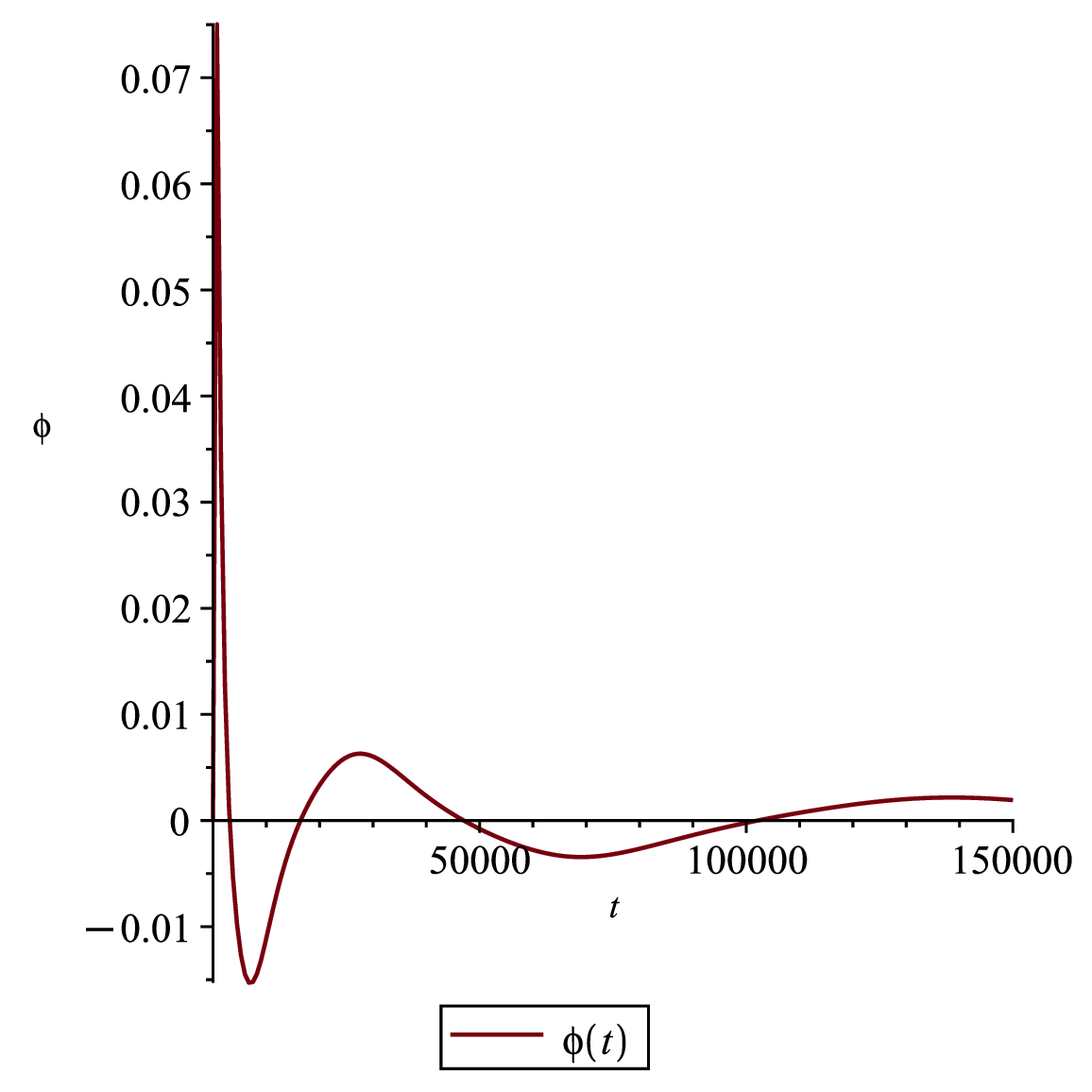}
  \end{subfigure}
  \hfill
  \begin{subfigure}[t]{.49\textwidth}
    \centering
    \includegraphics[width=\linewidth]{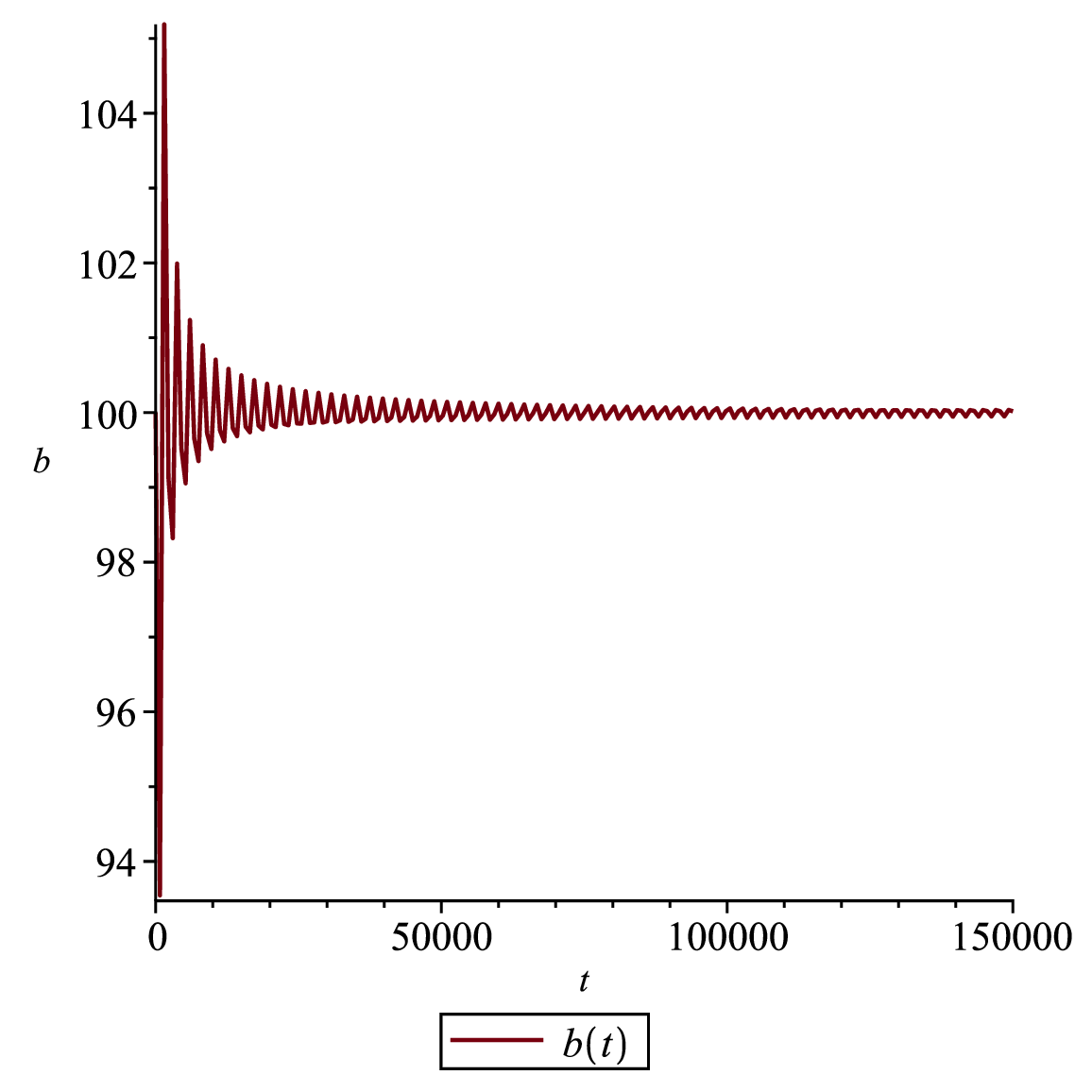}
  \end{subfigure}
  \caption{\footnotesize Oscillations of a) the scalar field $\phi(t)$ b) the scale factor $b(t)$ for $\alpha=2500,\; \Lambda=-0.0003,\; \xi=-10^{-12},\; \lambda=0.0001,\; \phi_0=2.45\cdot 10^{-6}$}
  \label{oscillations}
\end{figure}

From them we can see that large $b$ in Planck units needs large dimensionless $\alpha$, so that the case of small $\xi$ seems more physically natural.
An example is shown in Fig.~\ref{oscillations} where we see rapidly decaying oscillations of $b$ and prolonged slowly decaying oscillations of the scalar field.

As for the general case isotropic and singular solutions can also be possible outcomes of the dynamics. Our numerical studies indicate that for a wide range of $\lambda\in(10^{-12};10^{-1})$ compactification solutions disappear
with increasing $|\xi|$ (for $|\xi|$ being of the order of several units), and further, for $|\xi|$ being $\sim 100$, we see only singular solutions.

So that, the presence of spatially curved extra dimensions results in a drastic change in behaviour of the scalar field which acquires non-zero asymptotic value. We remind a reader that in the case of a perfect fluid or {\it minimally coupled} scalar field the only influence of extra dimensions after their stabilisation is to rescale the Newton and cosmological constants. All other features of background cosmological dynamics are the same as in 3+1 dimensional world without extra dimensions. On the contrary, dynamics of a {\it non-minimally coupled} field changes qualitatively and this change does not disappear after extra dimension stabilisation.

\section{Discussion}

In this paper we have considered the cosmological behaviour of a scalar field in Gauss-Bonnet gravity in the presence of spatially curved extra dimensions. We have found that if the scalar field is non-minimally coupled with the Ricci scalar $R$, the effects of extra dimensions do not decay after their stabilisation. Fixed point of the scalar field evolution shifts from the minimum of its potential to the minimum of a modified potential which acquires the additional massive term. Moreover, this term can be negative, turning thus usual $\lambda \phi^4$ potential of a self-interacting scalar field into a Mexican hat potential, needed for the Higgs mechanism to work.

It is worth to point out that compactification in Lovelock Gravity without scalar fields, once reached a regime with stabilized extra dimensions,
leads to a redefinition of effective Newton and cosmological constants in the large dimensions. However by coupling a scalar field non-minimally to gravity leads to a
 more dramatic effect as it qualitatively changes the shape of the potential of the scalar field
due to extra quadratic mass term.
 The quadratic mass term can be negative giving to the effective potential the shape of a Mexican hat and therefore leading to spontaneous symmetry breaking.
 The Mexican hat potential is the basic ingredient for Higgs field
to give masses to the fundamental particles of the standard model \cite{higgs1} , \cite{higgs2} , \cite{higgs3}, \cite{weinberg}. In the Higgs Mechanism the
 quartic term of the scalar potential must be positive in order to be bounded from below, however
the negative mass term cannot be justified from fundamental principles inside the framework of the standard model.

The idea of considering the massive term in Higgs potential as an effective one have been developed in several ways.  Gravity is involved in \cite{Odin3}, multidimenion set-up is used in \cite{Rubin}.
Our model is different in the sense that in \cite{Rubin}
 the Higgs field itself is an effective field, while in our paper scalar field with a quartic potential exists in a bare
Lagrangian, and the role of extra dimensions is to generate an extra term in the effective potential.

If we do not require zero $H$, the Mexical hat form can be got even without extra dimensions, as it was shown already in \cite{Vernov}. In this case we need a positive $\xi$ which gives a fixed point at $\phi^2=6\xi H^2/\lambda$. This rahter interesting case is beyond the scope of the present paper.

We can also note that resulting formula for the new scalar field fixed point does not contain the Gauss-Bonnet coupling constant $\alpha$. It happens  because the GB term does not directly modify the Klein-Gordon equation. Its role is only to stabilise extra dimensions while the last term in  (\ref{eq:scalar_field}) is responsible for the shifting of the scalar field fixed point. This means that any other mechanism for stabilisation, which does not modify Klein-Gordon equation is suitable as well. For example, presence of higher order Lovelock terms should not destroy the described picture. On the other hand, direct coupling of GB term with the scalar field can, in principle, change the situation. We leave detailed study of these problems to a future work.

\section*{ Acknowledgments} The work of AG is supported by grant FONDECYT 1240247.
 The work of AT  have been supported by  the Russian Government Program of Competitive Growth of Kazan Federal University. AT is grateful to Universidad
Austral de Chile, where part of this work have been done, for hospitality.

\end{document}